\documentstyle{article}
\begin{document}\large\rm
\baselineskip=0.79cm

\title{\bf Chaotic Simulated Annealing by A Neural Network Model with Transient Chaos}

\author{Luonan Chen \\{The Kaihatsu Computing Service Center, 2-2-18, Fukagawa, Koto-Ku, Tokyo 135, Japan} \and Kazuyuki Aihara \\{The University of Tokyo,
 7-3-1 Hongo, Bunkyo-Ku, Tokyo 113, Japan}}
\date{}
\maketitle
\clearpage
\setcounter{page}{1}
\pagestyle{headings}
This paper is published in {\bf Neural Networks,} Vol.8, No.6, pp.915-930,
1995. Note that 
the theoretical results related to this paper should be referred to 
"Chaos and Asymptotical Stability in Discrete-time Neural Networks" 
by L.Chen and K.Aihara, Physica D (in press).\\[1cm] 
Title:

\begin{center}\LARGE
{\bf Chaotic Simulated Annealing by A Neural Network Model with Transient Chaos}
\end{center}

\vspace*{1.0cm}

Authors:

\begin{center}
Luonan Chen\\
Power System Department,\\
The Kaihatsu Computing Service Center Ltd.,\\
2-2-18, Fukagawa,Koto-Ku, Tokyo 135, Japan\\
TEL: +81-3-3642-9771 \ \ \  FAX: +81-3-3642-9796
\end{center}

\vspace*{0.3cm}

\begin{center}
Kazuyuki Aihara\\
Department of Mathematical Engineering and Information Physics,\\
Faculty of Engineering,\\
The University of Tokyo,\\
7-3-1 Hongo, Bunkyo-Ku, Tokyo 113, Japan\\
TEL: +81-3-3812-2111(EXT.6911) \ \ \ FAX: +81-3-5689-5752
\end{center}

\vspace*{0.8cm}
Acknowledgments: The authors wish to express their thanks to S.Amari, Y. Akiyama, H.Ando, Y.Hayakawa, M.Kubo, I.Matsuba, T.Matsui,I.Tsuda and O.Watanabe for their valuable comments. This research was partially supported by a Grant-in-Aid for Scientific Research on Priority Areas from the Ministry of Education, Science and Culture of Japan.

\vspace*{0.3cm}
{\it Running title: A Transiently Chaotic Neural Net}

\clearpage  
\pagestyle{myheadings}
\markright{Abstract}

\begin{center}
{\bf Abstract}
\end{center}

We propose a neural network model with transient chaos, or  a transiently chaotic neural network (TCNN) as an approximation method for combinatorial optimization problems, by introducing 
transiently chaotic dynamics into neural networks. Unlike conventional neural networks only with point attractors, the proposed neural network has richer and more flexible dynamics,  so that it can be expected to have higher ability of searching  for globally optimal or near-optimal solutions.  A significant property of this model is that the chaotic neurodynamics is temporarily generated for 
searching and self-organizing, and eventually vanishes with autonomous decreasing  of a bifurcation parameter corresponding to the "temperature" in usual annealing process. Therefore, the neural network gradually approaches, through the transient chaos, to dynamical structure similar to such conventional models as the Hopfield neural network which converges to a stable equilibrium point.
 Since the optimization process  of the transiently chaotic neural network is similar to simulated  annealing, not in a stochastic way but in a deterministically chaotic way, the new method is regarded as  chaotic simulated annealing (CSA). Fundamental characteristics of the transiently chaotic neurodynamics are numerically investigated with examples of a single neuron model and the Traveling Salesman Problem (TSP). Moreover, 
 a  maintenance scheduling problem for generators in a practical power system is also analysed to verify practical efficiency of this new method.

\vspace*{5mm}

{\bf Keywords-} Neural network, Chaos, Transient chaos, Simulated annealing, TSP, Bifurcation, Combinatorial optimization problem, Modern heuristics, NP-hard

\clearpage
\pagestyle{myheadings}
\markright{List of Symbols}

\begin{center}
List of Symbols
\end{center}

\vspace*{2cm}

\begin{tabular}{ll}
$x_{i} $  &: output of neuron $i$ \\
$y_{i} $  &: internal state of neuron $i$ \\
$\omega_{ij}$  &: connection weight from neuron $j$ to neuron $i$\\
$I_{i}$   &: input bias of neuron $i$ \\
$\alpha$  &: positive scaling parameter for inputs\\
$k$       &: damping factor of nerve membrane ($0 \le k \le 1$)\\
$z_{i}, z $  &: self-feedback connection weight or refractory strength\\
$\epsilon$ &: steepness parameter of output function($\epsilon >0$)\\
$E$        &: energy function\\
$\Delta t$ &: time step of the Euler method \\
$\beta $   &: damping factor for $z_{i}$ and $z$\\
$\lambda $ &: the Lyapunov exponent\\
$I_{0}$  &: positive parameter\\
$G_{i}$  &: generating capacity of unit-$i$\\
$h$      &:total periods for the horizon\\
$mp_{i}$ &: maintenance periods of unit-$i$\\ 
$D_{j}$  &: load in period-$j$\\
$R_{j}$  &: reserve margin in period-$j$\\ 
$A_{l}$  &: set for units belonging to power plant-$l$
 \end{tabular}
 
$hf_{i} \sim ht_{i}$: specified maintenance periods for unit-$i$

\clearpage
\pagestyle{headings}

\section{Introduction}
A lot of problems in science and technology are related to combinatorial optimization problems. Researches in this field aim at developing efficient techniques for finding solutions realizing minimum or maximum values of a objective function subject to a set of constraints with many discrete variables.  A number of interesting combinatorial optimization problems are hard to deal with or intractable because they belong to a class of quite difficult problems, the NP (Nondeterministic polynomial) -hard class, which is a terminology of  computational complexity. That is, to find an exactly optimal solution for any NP-hard problem requires a number of computational steps that grows faster than any finite power of some appropriate measure of the problem size  as long as P $\neq$ NP. 

To address the NP issue in more details, we briefly give rough explanations of several terminologies. The  recognition or decision problems which have a known polynomial-time deterministic algorithm are said to be in class P. On the other hand, the decision problems solvable by a polynomial-time nondeterministic algorithm are in class NP, where $P \subseteq NP$ (Reeves, 1993). 
If every problem in NP is polynomially reducible to a problem Q, Q is NP-hard. Furthermore, if every problem in NP is polynomially transformable to a problem Q in NP, Q is NP-complete, which is a class of the hardest problems in NP.  Therefore, if a polynomial algorithm were to be found for any one of NP-hard or NP-complete problems, a polynomial algorithm would be also found for every problem in NP, i.e., P $=$ NP which means that all NP decision and NP optimization problems would have a polynomial algorithm (Karp, 1972; Garey and Johnson, 1979; Watanabe, 1994). 
A classical and famous example of the NP-hard problem is the optimization version of TSP (travelling salesman problem) whose decision version is NP-complete. 

To seek an optimal solution, a lot of exact methods have been so far proposed to deal with the combinatorial optimization problems, such as cutting plane methods, dynamic programming based on the Bellman's principle of optimality, branch-and-bound methods, and so on. 
 Unfortunately, no exact polynomial algorithm, however, has been found for any NP-hard problem up to now. In fact, it is widely believed  that P $\neq$ NP though not proved.

Since all attempts to look for an exact method with polynomial time for NP-hard problems have failed, many researchers have been paying attention, instead,  on approximate or heuristic algorithms  which seek near-optimal solutions at a reasonable computational cost without ensuring optimality or feasibility (Reeves, 1993). 
>From the viewpoint of the approximate algorithms, the concept of NP-hardness is further partitioned into weak NP-hardness and strong NP-hardness (Garey and Johnson, 1979; Ibaraki, 1994b). 
A typical problem with weak NP-hardness is the KNAPSACK problem, which has a fully polynomial-time approximation scheme (FPTAS) which gives a near-optimal solution with an arbitrary accuracy requirement $\epsilon (>0)$, by polynomial time of $1/ \epsilon $.
However, most of well-known NP-hard problems are not in the weak NP-hard class.  In fact, by the concept of Linear-reduction, MAX SNP (strict NP)- complete class and MAX SNP-hard class (Papadimitriou and Yanakakis, 1991) which includes many important NP-hard problems, can be defined. 
It has been proved (Arora, Lund, Motwani, and Szegedy, 1992; Ibaraki, 1994b) that there exists no polynomial-time approximation schemes which can give a solution  arbitrarily close to the exact one for MAX SNP-hard class if P $\neq$ NP. Moreover, for some NP-hard problems not belonging to MAX SNP-hard class, such as TSP without triangle inequality of distances,  even an approximate algorithm with guarantee of a finite accuracy does not exist as long as P $\neq$ NP. 

Nevertheless, from the practical viewpoint, approximate algorithms are useful and have achieved considerable success 
when applied to a large number of problems.
The approximate algorithms developed so far include greedy methods, local search methods, randomized methods, relaxation techniques, partial enumeration, decomposition and partition approaches, and so on (Zanakis, Evans and Vazacopoulos, 1989; Reeves, 1993). Among them, the local search methods have been intensively studied, where a concept of PLS (polynomial time local search) - complete is also introduced on the basis of PLS- reducibility to evaluate the computational complexity (Johnson, Papadimitriou, \& Yannakakis, 1988). Moreover, 
 some modern heuristic or meta heuristic algorithms which are applicable more generally to combinatorial optimization problems rather than problem-specific (Reeves, 1993) and which can be related to physical and biological phenomena (Bounds, 1987), have recently been developed on the basis of local search; they include  simulated annealing, tabu search, genetic algorithms and artificial neural networks (Reeves,1993; Ibaraki, 1994a).
  In this paper, we focus on two of such modern heuristics, i.e., simulated annealing and neural networks as well as chaotic dynamics in order to develop a new heuristic technique, which we call chaotic simulated annealing for combinatorial optimization problems. Since simulated annealing and neural networks are our main topics, we give a brief review next.

The main difficulty in solving combinatorial optimization problems is that straightforward algorithms, such as greedy methods and local search methods, tend to become trapped in local minima.
To avoid getting stuck in local minima,
in 1983 Kirkpatrick et al. developed simulated annealing with  the Monte Carlo technique first proposed by Metropolis et al.(1953), which emulates the solid annealing processing by first heating a solid to its melting point and then slowly cooling the material at a natural rate related to its heat transport speed (Kirkpatrick et al, 1983). Since the optimization processing is usually undertaken in a stochastic manner, this annealing mechanics is actually  {\it stochastic  simulated annealing (SSA)}. 
Convergence in distribution with the uniform measure over the exact solutions, or globally optimal solutions has been theoretically proved by Geman and Geman as long as the cooling schedule is slow enough to be at least inversely proportional to the logarithm of time (Geman \& Geman, 1984). 
A fast annealing with generating probability obeying the Cauchy distribution, which cooling scheduling is inversely linear in time (Szu \& Hartley, 1987), and a very fast annealing with exponential cooling schedule (Ingber \& Rosen, 1992; Ingber, 1993) have been developed 
to improve the performance of the original SSA (Rosen \& Nakano, 1994).

Up to now, the SSA method 
 has been widely applied to various optimization problems with considerable success, but nevertheless it still suffers from several shortcomings.
  In particular, the method requires subtle adjustment of  parameters in the annealing schedule such as   the size of the 
temperature steps during annealing, the temperature range, the number of re-starts and re-direction of the search,  etc., due to the  intrinsic characteristics of its stochastic mechanics (see Kirkpatrick, Gelatt Jr., \& Vecchi, 1983; Aarts \& Korst, 1989; Johnson et al., 1989; Johnson et al., 1991; Ingber, 1993; Reeves, 1993; Rosen \& Nakano, 1994). In addition, time consuming due to its Monte Carlo scheme is also an annonying problem of SSA. That is, to guarantee convergence to an exact solution, SSA would requires even more iterations than complete enumeration for some problems. For example, for TSP with $n$ cities,  the computation cost of SSA is $O(n^{n^{2n-1}})$ far more than $O((n-1)!)$ of complete enumeration and 
$O(n^{2}2^{n})$ of dynamic programming, when the logarithmic cooling schedule is used (Aarts \& Korst, 1989; Kubo, 1994). Therefore, heuristic fast cooling schedules,
 such as geometric cooling (Johnson, Aragon, McGeoch, and Schevon, 1989, 1991), exponential cooling  and adaptive cooling (Ingber, 1993; Reeves, 1993), are applied to SSA to improve its performance.

On the other hand, as one of meta heuristics, neural network approaches have also been shown to be a  powerful tool for combinatorial optimization (Hopfield \& Tank, 1985; Rumelhart, et al., 1986; Wells, 1992; Peterson \& Soderberg, 1993). One of the typical models of artificial neural networks is the Hopfield neural networks with symmetric connection weights (Hopfield, 1982, 1984; Hopfield \& Tank, 1985).
         Although these neural networks guarantee convergence to a stable equilibrium point due to their gradient descent dynamics, the main drawback is that these neural networks suffer from the local minimum problems whenever applied to optimization problems. Therefore, various simulated annealing techniques have been combined with neural networks to overcome the weakness. A typical example of such neural networks is Boltzmann machines composed of stochastic binary neurons with firing probability of the logistic function (Ackey et al., 1985; Aarts \& Korst, 1989). Besides, Akiyama et al. (1991) developed  Gaussian Machines by adding  random noises with the Gaussian distribution to an analog neural network, to perform SSA processing. 
         On the other hand, instead of stochastic models, several heuristic annealing approaches which are actually deterministic simulated annealing (DSA), have been proposed and applied to neural networks; the approaches include gain sharpening of the Hopfield networks (Hopfield \& Tank, 1985; Lee \& Sheu, 1991; Akiyama et al., 1991),  and mean field approximation annealing (Peterson \& Anderson, 1987; Hinton, 1989; Geiger \& Girosi, 1991, Masui \& Matsuba, 1991; Peterson \& Soderberg, 1993), and so on.

While most of the conventional networks fundamentally utilize gradient descent dynamics, a number of artificial neural networks with chaotic dynamics have been investigated towards more complex neurodynamics (e.g., Sompolinsky, Crisanti \& Sommers, 1988; Aihara, Takabe \& Toyoda, 1990; Aihara, 1990; Yao \& Freeman, 1990; Tani, 1991; Inoue \& Nagayoshi, 1991; Tsuda, 1992; Nozawa, 1992; Chen, 1993;  Adachi et al., 1993; Yamada et al., 1993; Nara, Davis \& Totsuji, 1993). Among them, we take in this paper
{\it chaotic neural network (CNN)} models which have been proposed by considering graded responses, relative refractoriness and spatio-temporal summation of inputs (Aihara et al., 1990; Aihara, 1990). Unlike the static or near-equilibrium behavior of the conventional neural networks, CNN has 
 richer and far-from-equilibrium dynamics with various coexisting attractors, not only of fixed points and periodic points but also of strange attractors in spite of the simple equations.
Although many reports have approved that the chaotic dynamics can be  promising techniques for information processing and optimization, the convergence problems have not yet been satisfactorily solved in relation to chaotic dynamics as far as the authors know.  That is, it is usually difficult to decide when to terminate the chaotic dynamics, or how to control chaotic behavior for convergence to a stable equilibrium point.  

In order to make use of the advantages of both the chaotic neurodynamics and the conventional convergent neurodynamics,  this paper aims at 
  developing a {\it transiently chaotic neural network (TCNN)} 
  for combinatorial  optimization problems, by introducing a new variable corresponding to the "{\it temperature}" in usual annealing process  into the CNN models to harness the chaotic dynamics appropriately.  
 Different from the convergent mechanism  of the Hopfield neural networks with gradient descent dynamics,
 the proposed model has richer and more flexible dynamics, which can prevent the network from being trapped in  commonplace local
  minima, and may obtain acceptably near-optimal solutions eventually. 
    A significant property of the new method is that the chaotic 
    neurodynamics which is temporarily generated for  searching and self-organizing,  gradually vanishes with autonomous decreasing of a bifurcation parameter, or the "temperature", accompanied with successive bifurcations from strange attractors into an equilibrium point. In other words,
   the neural network gradually approaches to dynamical structure similar to  the Hopfield neural networks, and usually converges to a stable equilibrium solution after vanishing of temporary chaotic behavior 
   which we call "transient chaos" (Chen, 1993).  
      Since the optimization process of the transiently chaotic neural network
        is deterministically chaotic  rather than stochastic,
       the proposed approach is regarded as {\it chaotic simulated annealing (CSA)},   in contrast to the conventional stochastic simulated annealing (SSA), from the viewpoint of annealing mechanics. 
   There are mainly two significant differences between CSA and SSA. The first  is that  SSA is {\bf stochastic} on the basis of the Monte Carlo scheme while  CSA is {\bf deterministic} with transiently chaotic dynamics. The second is that the convergent processing of SSA is undertaken by control of stochastic "thermal" {\bf fluctuations} while  that of CSA is  by control of {\bf bifurcation} structures. Furthermore, the searching  space for optimization is also different; namely,  SSA fundamentally searches all possible states by temporally changing probability distributions achieved by its Monte Carlo scheme; on the other hand, the searching of CSA with temporally changing invariant  measures determined by its dynamics is restricted to a possibly fractal subspace with continuous states.
      Since the dynamics of TCNN is dissipative, the volume of the searching region in CSA is usually very small compared with that of the state space. 
 Therefore, CSA can be expected to  perform  efficient searching because of its reduced search spaces if the restriction is adequate to include a global optimum state or its good approximation or  at least some  parts of the basins for such solutions. Since the dynamics of TCNN is also deterministic, CSA can be viewed as  a method belonging to {\it deterministic simulated annealing (DSA)}.
 
   To verify and test the new method,  transiently chaotic dynamics of a single neuron model and networks for  the TSP (Traveling Salesman Problem) is numerically    analysed in this paper.  Moreover,  a maintenance scheduling problem for generators in a power system is also studied to certify the ability of TCNN with CSA as a method of modern heuristics    for practical combinatorial problems.                           
   
This paper is organized as follows:  In the next section, the CNN model is briefly described, and then a transiently chaotic neural network as well as the chaotic simulated annealing is proposed for combinatorial optimization problems, and numerically analysed  to clarify the nonlinear computational dynamics.
 In section three,
 TSP with 4 cities, 10 cities and 48 cities are analysed to investigate efficiency of the  model.  A large-scale combinatorial optimization problem for a practical generator maintenance scheduling of a power system is examined in section four. Finally, we give some general remarks to conclude this paper.

\clearpage
\pagestyle{myheadings}
\markright{2 TRANSIENTLY CHAOTIC NEURAL NETWORKS AND CSA}

\section{Transiently Chaotic Neural Networks and Chaotic Simulated Annealing}

In this section, firstly the chaotic neural network (CNN) is briefly summarized. Then, a transiently chaotic neural network (TCNN) is introduced. In the  last of this section, a single neuron model is analysed to demonstrate both transiently chaotic dynamics and chaotic simulated annealing (CSA).

\subsection{A model of the chaotic neural network}

The model of the chaotic neural network (CNN) can be described as follows for the purpose of this paper (Aihara, Takabe, \& Toyoda, 1990):

\begin{eqnarray}
x_{i}(t) = \frac{1}{1+e^{- y_{i}(t) / \epsilon}}  \label{e1}\\
y_{i}(t+1) = k y_{i}(t) + \alpha (\sum_{j=1}^{n} \omega_{ij}x_{j}(t) + I_{i}) - z_{i} x_{i}(t) \label{e2}\\
\ \ \ \ (i=1,...,n) \nonumber 
\end{eqnarray}

where, 

\begin{tabular}{ll}
$x_{i} $  &: output of neuron $i$\\
$y_{i} $  &: internal state of neuron $i$\\
$\omega_{ij}$  &: connection weight from neuron $j$ to neuron $i$\\
$I_{i}$   &: input bias of neuron $i$\\
$\alpha$  &: positive scaling parameter for inputs\\
$k$       &: damping factor of nerve membrane($0 \le k \le 1$)\\
$z_{i}$  &: self-feedback connection weight or refractory strength (constant) ($z_{i}>0$)\\
$\epsilon$ &: steepness parameter of the output function ($\epsilon >0$).
\end{tabular}

Unlike simple threshold elements in conventional neural networks with gradient descent dynamics converging to an equilibrium point, the neuron model of CNN has  more complex dynamics as a constituent element of neural networks and produces  rich spatio-temporal dynamics as networks (Aihara, et al., 1990; Aihara 1990; Adachi et al., 1993; Yamada et al., 1993).

\subsection{A model of a transiently chaotic neural network}

As already mentioned, such conventional neural networks as the Hopfield neural networks with continuous-time or asynchronously discrete-time state transitions, guarantee convergence to a stable equilibrium solution but suffer from local minimum problems.  
On the other hand, although the dynamics of CNN has  an intriguing property to move chaotically over fractal structure in the phase space, without getting stuck at local minima because of accumulation of refractory or self-inhibitory effects (Aihara et al., 1990; Aihara, 1990; Nozawa, 1992; Adachi et al., 1993; Yamada, et al., 1993), the convergence problems of the chaotic dynamics have not been satisfactorily solved so far.  

In order to take advantages of both the convergent dynamics and the chaotic dynamics but to overcome their disadvantages for combinatorial optimization problems,  we propose a transiently chaotic neural network (TCNN)  by modifying the chaotic neural network, as defined below: 
 
\begin{eqnarray}
x_{i}(t) = \frac{1}{1+e^{-  y_{i}(t) / \epsilon }}  \label{e3}\\
y_{i}(t+1) = k y_{i}(t) + \alpha(\sum_{j=1,j \ne i}^{n} \omega_{ij}x_{j}(t) + I_{i}) - z_{i}(t)(x_{i}(t)-I_{0}) 
\label{e4} \\
z_{i}(t+1)=(1-\beta)z_{i}(t)  \label{e5}\\ 
 (i=1,...,n) \nonumber
 \end{eqnarray}
 
where, 
$\omega_{ij}= \omega_{ji}; \omega_{ii}=0$; $\sum_{j=1,j \ne i}^{n}
\omega_{ij}x_{j}+I_{i}=-\partial E/\partial x_{i}$

\begin{tabular}{ll}
$E$ &: energy function defined in section 3\\
$z_{i}(t)$  &: self-feedback connection weight or  refractory strength ($z_{i}(t) \ge 0$)\\
$ \beta$ &: damping factor of the time-dependent $z_{i}(t)$ ($0 \le \beta \le 1$)\\
$I_{0}$ &: positive parameter.
\end{tabular}
 
The difference between CNN and TCNN is the third terms in the right-hand sides of eqn.(\ref{e2}) and eqn.(\ref{e4}) where $z_{i}x_{i}(t)$ in eqn.(\ref{e2}) is replaced with  $z_{i}(t)(x_{i}(t)- I_{0})$ in eqn.(\ref{e4}).  The  term  can be related to  negative (inhibitory) self-feedback or refractoriness (Aihara et al., 1990) with a bias $I_{0}$.  We will show later that this neural network has actually  transiently chaotic dynamics which eventually converges to a stable equilibrium point through successive bifurcations like a route of reversed period-doubling bifurcations, with the temporal evolution of  a new variable $z_{i}(t)$ according to eqn.(\ref{e5}). The variable $z_{i}(t)$ corresponds to the temperature in usual stochastic annealing process. Thus, eqn.(\ref{e5}) represents an exponential cooling schedule for the annealing.
   
In order to compare TCNN with the continuous-time Hopfield neural network (Hopfield, 1984; Hopfield \& Tank, 1985), eqns.(\ref{e3})-(\ref{e5}) can be rewritten in forms of differential equations as follows:                                                
 \begin{eqnarray}
x_{i}=\frac{1}{1+e^{- y_{i} / \epsilon }}  \label{e6}    \\
\frac{dy_{i}}{dt} = - k' y_{i} + \alpha ' (\sum_{j=1}^{n} \omega_{ij}x_{j} + I_{i}) - z_{i}(x_{i}-I_{0})  \label{e7} \\
\frac{dz_{i}}{dt} = -\beta ' z_{i}   \label{e8} \\
(i=1,...,n) \nonumber                                             
\end{eqnarray}

where, $\omega_{ij}= \omega_{ji}; \omega_{ii}=0$; $\sum_{j=1,j \ne i}^{n} \omega_{ij}x_{j}+I_{i}=-\partial E/\partial x_{i}$. 
Obviously, eqns.(\ref{e3})-(\ref{e5}) are the difference equational version of eqns.(\ref{e6})-(\ref{e8}) on the basis of the Euler method.
Comparing eqns.(\ref{e3})-(\ref{e5}) with eqns.(\ref{e6})-(\ref{e8}),  we get the relationships of $(1- \Delta t k')=k$, $\Delta t \alpha ' = \alpha$, and  $\Delta t \beta ' = \beta$, where $\Delta t$ is the time step by the Euler method.

The difference between the proposed model and the continuous-time Hopfield neural network
 is that a nonlinear term $-z_{i}(x_{i}- I_{0})$ is added in eqn.(\ref{e7}). Since the "temperature" $z_{i}(t)$  tends   toward zero with time evolution in the form of $z_{i}(t)=z_{i}(0)e^{- \beta ' t}$, eqns.(\ref{e6})-(\ref{e8}) eventually reduce
  to the continuous-time Hopfield neural network without self-feedback connections, namely  $w_{ii} =0$. From the observation of eqn.(\ref{e7}), as the variable $z_{i}$
 can be interpreted as the strength of negative (inhibitory) self-feedback connection of each neuron,  {\bf the annealing mechanics is regarded as  exponential  damping of the negative self-feedback connection weights}.
 
 In addition, the relation between the added term and $z_{i}$ implies that the dynamics of this model depend sensitively on  $z_{i}$. In fact, if the value of $z_{i}$ is fixed, the model of eqns.(\ref{e6})-(\ref{e8}) is equivalent to CNN with complicated bifurcation structure (Aihara, et al., 1990; Aihara, 1990) by difference analogue with the Euler method and  variable transformation (Nozawa, 1994). The damping of $z_{i}$  produces successive bifurcations in the difference version so that the neurodynamics eventually converges from strange attractors to a stable equilibrium point as will be shown later.
   In this paper,  for the sake of simplicity,  $z_{i}$ is set to decay  exponentially as described by eqn.(\ref{e8}) or eqn.(\ref{e5}), although more skillful damping may be required for more efficient optimization processing.

\subsection{Transiently chaotic dynamics of the single neuron model}

We examine nonlinear dynamics of the single neuron model in this section. 

>From eqns.(\ref{e6})-(\ref{e8}), the single neuron model is derived by the Euler method as follows (or directly from eqns.(\ref{e3})-(\ref{e5})): 

\begin{eqnarray}
x(t) =\frac{1}{1+e^{-  y(t)/ \epsilon }}  \label{e9}\\
y(t+1) =ky(t)+ \gamma -z(t)(x(t)-I_{0})  \label{e10}\\
z(t+1) =(1-\beta )z(t)  \label{e11} 
\end{eqnarray}

where, $\gamma =\alpha I_{i}$. Obviously, if the value of $z(t)$ is fixed at $z_{0}$, eqns.(\ref{e9})-(\ref{e10}) can be easily transformed to the following  one-dimensional mapping from $y(t)$ to $y(t+1)$ by substituting eqn.(\ref{e9}) into eqn.(\ref{e10}): 

\begin{equation}
y(t+1) =ky(t)+ \gamma -z_{0}(\frac{1}{1+e^{-  y(t)/ \epsilon }}- I_{0}) \ \ .    \label{e91}
\end{equation}

The values of the parameters in eqns.(\ref{e9})-(\ref{e11}) are set as follows:

\begin{equation}
 k=0.9; \epsilon = 1/250; I_{0}=0.65;  z(0)=0.08
\label{e12}
\end{equation}

In the following,  we vary only $\beta$ and $\gamma$ (or $\alpha$) to investigate  the dynamics while other parameters are fixed as eqn.(\ref{e12}).

Fig.\ref{f1} shows the time evolutions of $x(t),z(t)$ and the Lyapunov exponent  $\lambda$ of  $x(t)$, which are calculated by eqns.(\ref{e9})-(\ref{e11}) with  $\beta =0.001, \gamma =0.0$ and the initial condition $y(0)=0.5$. Here, the Lyapunov exponent $\lambda$ which is generally taken as a crucial index to identify orbital instability of deterministic chaos, is defined as follows:

\begin{equation}
\lambda = \mathop{lim}_{m \rightarrow + \infty} \frac{1}{m} \sum_{k=0}^{m-1} ln|\frac{dx(k+1)}{dx(k)}| \ \ \ \ \ \ .
  \label{e95}
\end{equation}
 
The  Lyapunov exponent is calculated with eqns.(\ref{e9}),(\ref{e91}) and (\ref{e95}) for sufficient large iterations $m$ for each fixed value of $z_{0}$. Namely, positive values of $\lambda$ indicate that dynamics of eqn.(\ref{e91}), or eqns.(\ref{e9})-(\ref{e10}) actually has orbital instability for the corresponding fixed value of $z_{0}$.

On the other hand, the time evolutions of $x(t)$ and $z(t)$ in Fig.\ref{f1} are calculated by simultaneous simulation with all of eqns.(\ref{e9})-(\ref{e11}). 
 Fig.\ref{f1} shows that, with exponential damping of  $z(t)$, the neuron output $x(t)$ gradually transits from chaotic behavior to a fixed point through  {\it reversed} period-doubling bifurcations; that is,  $x(t)$ behaves erratically and unpredictably during the first 300 iterations and eventually converges to a stable fixed point around the iteration 950 through e.g. the reversed period-doubling bifurcations.  Since the Lyapunov exponents in the first 300 iterations are mostly positive, the behavior is understood as chaotic in most of the region with possible periodic windows. Fig.\ref{f1} indicates that 
  chaotic fluctuations decrease with the damping of $z(t)$ and eventually vanish.  In other words, it is a kind of  transiently chaotic dynamics, 
and the dynamical structure of the neural network almost coincides with the Hopfield network, when the value of $z(t)$ decreases small enough.

Furthermore, we vary the values of the parameters $\beta$ and $\gamma$, to examine their influences on the neural dynamics. Fig.\ref{f2}(a) and Fig.\ref{f2}(b) are the time evolutions of $x(t)$ and $z(t)$, when $\beta =0.002$ and $\beta =0.0008$, respectively.
 In Fig.\ref{f2}(a), the chaotic dynamics of $x(t)$ vanishes quickly because $z(t)$ decreases rapidly owing to the larger value of $\beta$.  On the other hand, in Fig.\ref{f2}(b), the chaotic dynamics of $x(t)$ in turn lasts longer due to the small value of $\beta$. Therefore, it is evident that $\beta$ governs the bifurcation speed of the transient chaos.

The convergence procedure of the transiently chaotic neural network is  fully deterministic, namely, starting  from deterministically chaotic dynamics,  through a reversed period-doubling route with
decreasing the value of $z$ which corresponds to the temperature of usual annealing, finally reaching a stable equilibrium solution. 
 The mechanics of TCNN is called {\it chaotic simulated annealing (CSA)} which is regarded as a kind of {\it deterministic simulated annealing (DSA)}.  Moreover, compared with the stochastic thermal fluctuations of SSA, the convergence process of CSA is characterized as  the nonlinear bifurcations which drive the neurodynamics to converge from strange attractors to an attracting fixed point.

\clearpage
\pagestyle{headings}

\section{Application to the Traveling Salesman Problem}

A classical and famous combinatorial optimization problem which is NP-hard is the travelling salesman problem  (TSP) which  seeks the shortest route through $n$ cities, visiting each once and only once and returning to the starting point. 
 TSP can be generally divided into symmetric TSP with $d_{ij} = d_{ji}$, and asymmetric TSP with $d_{ij} \neq d_{ji}$, where $d_{ij}$ is the distance from city-$i$ to city-$j$. For symmetric TSP, there are $(n-1)!/2$ possible tours (or solutions), while there are $(n-1)!$ possible tours for asymmetric TSP. Obviously, the number of possible tours grows faster than any finite power of $n$. 
As opposed to intuition, asymmetric TSP is actually easier to solve than symmetric TSP in cases such that distances are randomly generated(Kubo,1994). 
Moreover, to categorize TSP further, it is important whether the TSP satisfies triangle inequality of distances, $d_{ik} + d_{kj} \geq d_{ij}$ or not. If a symmetric TSP satisfies the triangle inequality, there exists an approximate algorithm which can give a route whose length is less than $3/2$ times of that of the exact solution (Christofides, 1976; Garey \& Johnson, 1979; Kubo, 1994). A typical example of the symmetric TSP satisfying the triangle inequality is TSP on the plane under the Euclidean metric.

TSP is simple to state but difficult to solve exactly due to the NP-hardness. For example, at a reasonable computational cost, it is thought that both usual integer programming and dynamic programming can deal only with 20 $\sim$ 30 city TSP (Ibaraki, 1994a). Recently, it was reported that 2392 and 4461 city symmetric TSP had been solved by branch and cut methods (Padberg, \& Rinaldi, 1991; Ibaraki,1994a).  
Compared with symmetric TSP, as mentioned above,  asymmetric TSP is 
 easier to deal with. In fact, a 500000 city asymmetric TSP has been solved by the branch-and-bound method (Miller \& Pekny, 1991).                
       
On the other hand, to find acceptably near-optimal solutions of TSP,  a large number of approximate algorithms have been developed, such as, greedy methods, nearest neighbor methods, insertion methods, the Christofides' algorithm, space filling curve methods, the Lin-Kernighan algorithm, 2-OPT, 3-OPT and so on (Kubo, 1994).  Among them, the Lin-Kernighan algorithm (Lin \& Kernighan, 1973) is well known to be very efficient in both the speed and the quality of solutions found; it is reported that solutions within 2\% of the optimum for symmetric TSP on the plane with as many as 50000 cities are found by the  Lin-Kernighan algorithm (Johnson, 1987; Johnson et al., 1988).

Since Hopfield and Tank (1985) applied their neural networks to TSP, TSP has been intensively studied in the field of artificial neurocomputing. It was recently reported (Nozawa, 1992; Yamada et al., 1993; Nozawa, 1994) that CNN or equivalent models may have efficient searching ability for such combinatorial optimization problems as TSP because they can escape from local minima due to accumulated  negative self-feedback inputs or refractory effects (Aihara et al., 1990). This searching ability of CNN, however, is a two-edged sword because CNN frequently escapes even from globally optimal or near-optimal states. Therefore, it is an important problem to make CNN converge to good solutions while utilizing the searching ability transiently. This is the problem to which TCNN with CSA may be useful.

In this paper, symmetric TSP on the plane is used to test the ability of TCNN with CSA as a heuristic method for combinatorial optimization. 
We adopt the formulation for TSP  by Hopfield and Tank (1985). Namely, 
 a solution of TSP with  $n$ cities is represented by a 
$n\times n$ - permutation matrix, where each element corresponds to output of a neuron in a network with  $n \times n$ lattice structure. Assume  $x_{ij}$ to be the neuron output which represents to visit city-$i$ in visiting order-$j$.  A computational energy function which is to minimize the total tour length while simultaneously satisfying all constraints takes the following form:

\begin{eqnarray}
E=\frac{W_{1}}{2}\{ \sum_{i=1}^{n} (\sum_{j=1}^{n}x_{ij}-1)^{2} + \sum_{j=1}^{n} (\sum_{i=1}^{n}x_{ij}-1)^{2} \}  \nonumber \\
+ \frac{W_{2}}{2} \sum_{i=1}^{n} \sum_{j=1}^{n} \sum_{k=1}^{n} (x_{kj+1}+x_{kj-1})x_{ij}d_{ik} 
\label{e15}
\end{eqnarray}

where  $x_{i0}=x_{in}$ and $x_{in+1}=x_{i1}$. $W_{1}$ and $W_{2}$ are the coupling parameters corresponding to the constraints and  the cost function of the tour length, respectively, where $d_{ij}$ is the distance between city-$i$ and city-$j$.  

By setting each connection weight $\omega_{ij}$ as same as the Hopfield neural network, the difference equations describing  the network dynamics of TCNN for the TSP are obtained from eqns.(\ref{e3})-(\ref{e5}) as follows:

\begin{eqnarray}
x_{ij}(t)=\frac{1}{1+e^{- y_{ij}(t)/ \epsilon }}  \label{e16}\\
y_{ij}(t+1)=ky_{ij}(t) - z(t)(x_{ij}(t)-I_{0}) 
 + \alpha \{-W_{1} (\sum_{l \ne j}^{n} x_{il}(t) + \sum_{k \ne i}^{n} x_{kj}(t)) \nonumber \\
  -W_{2}(\sum_{k \ne i}^{n} d_{ik}x_{kj+1}(t) + \sum_{k \ne i}^{n}d_{ik}x_{kj-1}(t) ) + W_{1}\}   \label{e17} \\
  (i,j=1,...,n) \nonumber\\
z(t+1) = (1- \beta ) z(t)   \label{e18}  
\end{eqnarray}

where the variable $z(t)$ is assumed to be common to all the neurons, and $I_{ij}=W_{1}$.

In order to improve the computation performance,  we code a discrete neuron output $x_{ij}^{D}$ instead of the continuous output $x_{ij} \ (i,j=1,...,n)$ as follows:

\begin{equation}
x_{ij}^{D}(t)= \left\{
\begin{array}{ll}
1 & \mbox{iff} \ \ x_{ij}(t)>\sum_{k=1}^{n}\sum_{l=1}^{n}x_{kl}(t)/(n\times n) \\
0 & \mbox{otherwise} 
\end{array}
\right.
\label{e19}
\end{equation} 

Since the output $x_{ij}$ takes a continuous value between 0 and 1, the corresponding energy function of eqn.(\ref{e15}) is also continuous.  We call the energy function of eqn.(\ref{e15}) with the continuous output $x_{ij}$ and that with the discrete output 
$x_{ij}^{D}$ as the {\bf Continuous Energy Function} $E^{C}$ and the
{\bf Discrete Energy Function} $E^{D}$, respectively.

In the following simulations,  $W_{1}=W_{2}=1$ and other parameters  are set as eqn.(\ref{e12}) except $\alpha$ and $\beta$, and states of neurons are cyclically  updated.

\subsection{Experiments on TSP with 4 cities}

Fig.\ref{f4} shows the  Hopfield-Tank original data on TSP with 10 cities (Wilson \& Pawley, 1988).  In this subsection, we use only the {\bf first 4 cities} to examine fundamental characteristics of TCNN when applied to TSP.

Fig.\ref{f5} shows the time evolutions of the continuous and discrete energy functions and $z$, when $\alpha =0.015$ and $\beta =0.001$.
Fig.\ref{f6} illustrates the time evolution of each neuron output $x_{ij}$, while Fig.\ref{f7} is amplification of $x_{11}$.

 Figs.\ref{f5}-\ref{f7} shows that the time evolution of the TCNN changes from chaotic behavior with larger fluctuations 
 at the early stage, through  periodic behavior with the period 5 and chaotic behavior with smaller fluctuations, 
 into the later convergent stage. 
   After about iteration 1300, when $z$ becomes so small that the convergent characteristic dominates the dynamics,  
   the neural network state finally converges to a fixed point corresponding to a global minimum (tour length = 1.18).
 In Figs.\ref{f5}-\ref{f7} and the subsequent simulations, one iteration means one cyclic updating of all neuron states. The output patterns in Fig.\ref{f6} indicates that the best route  is 
3-4-1-2-3.  In addition, it is found that other representations of best routes with the same energy value (tour length = 1.18) such as 2-3-4-1-2 and 4-1-2-3-4
can be obtained, depending on the initial conditions of $y_{ij}$.

The simulation results imply that, 
 TCNN firstly reaches a valley  with noninteger solutions probably involving a part of the basin of a certain global minimum after $z$ becomes small enough for disappearance of  transient chaos, and then TCNN converges to the global minimum with an integer solution starting from the state in the valley with further decreasing of $z$. 
For instance, according to Figs.\ref{f5}-\ref{f7}, 
at about iteration 1100, the system reaches a valley with a noninteger solution  after the transiently chaotic behavior, and then  rapidly transits from the {\it noninteger solution} through falls toward a nearly {\it integer solution} which corresponds to the global minimum.
 Consequently, the dynamics of the TCNN can be classified  into two phases, depending on the value of $z$, 
i.e., (1) the  "{\bf chaotic searching phase}" with  transiently chaotic dynamics for  a relatively large $z$,
 and (2) the "{\bf stable convergent phase}" with convergent dynamics similar to the Hopfield neural network for a relatively small $z$. 
 The chaotic searching phase seems to act as searching for a valley belonging to the basin of a certain global minimum, while the stable convergent phase, starting from the valley, is to press the neural network to converge to the corresponding integer solution of the global minimum.
 For the case of Figs.\ref{f5}-\ref{f7}, the  "chaotic searching phase" is until about iteration 1100, and the "stable convergent phase" starts at about iteration 1100 when $z$ is small enough.

 Moreover, it is also found that the period-5 behavior in Figs.\ref{f5}-\ref{f7} is actually incomplete {\it traveling waves} (Wells, 1992), where the solutions correspond to local minima.

Figs.\ref{f8}-\ref{f9} show the time evolutions of the continuous energy function when $\alpha$ and $\beta$ are largely varied. 
For the cases of Fig.\ref{f8}(b) and Fig.\ref{f9}(b) where
 $\beta$ and  $\alpha$ is extremely large respectively, the chaotic dynamics vanishes too quick to realize sufficient chaotic searching.  Therefore, the solution rapidly converges to a local minimum like  the Hopfield neural network.  In contrast, for the case of Fig.\ref{f8}(a) where $\beta$ is very small, the strong chaotic dynamics lasts too long to converge within the prescribed iterations.  On the other hand, for the case of Fig.\ref{f9}(a) where 
  $\alpha$ is extremely small so that the energy function can not sufficiently affect 
the neurodynamics, the network converges to a wrong solution.  The obtained solutions in these cases are either infeasible solutions or local minima.

It should be noticed that $\alpha$ represents the influence of the energy function to the neurodynamics, or the balance between the self-feedback or refractory term inducing the chaotic dynamics and the gradient term $\partial E/\partial x_{ij}$ inducing the convergent dynamics.
If $\alpha$ is too large, the influence of the energy function becomes too strong to generate the transient chaos. On the other hand, the energy function cannot be sufficiently reflected in the neurodynamics if $\alpha$ is too small.

\subsection{Solution for TSP with 10 cities and 48 cities}

As a next example, we have analysed 10 city TSP of the data in Fig.\ref{f4} with $\alpha =0.015$, $\beta =0.015,0.010,0.005,0.003$, and with  5000 different initial conditions of $y_{ij}$ generated randomly in the region [-1,1] for each value of $\beta$. The results are summarized
in Table \ref{t1}.

As shown in Table \ref{t1}, TCNN  converges averagely with 81 iterations (for  discrete energy function) when $\beta =0.015$, where among 5000 cases, 31 cases are to local minima (tour length = 2.74,2.78,...), 23 cases are to infeasible solutions and 4946 cases are to global minima (Tour length=2.70).  That is, the rate of global optimization is 98.9\% for $\beta =0.015$.  The best route is shown in Fig.\ref{f4}. 
As mentioned previously, the different initial conditions of $y_{ij}$ can derive  different output representations of global optimal solutions with the same tour length.

The parameter $\beta$ can be considered as a damping speed parameter of the negative self-feedback strength or the "temperature" $z$, which controls the annealing schedule.
In order to produce stronger and longer chaotic dynamics,
 we set $\beta$ small as $\beta = 0.005$ and $0.003$, so that the temperature $z$ decreases more slowly. According to Table \ref{t1}, the results are considerably improved, e.g., when $\beta =0.005$, TCNN 
   converges averagely with 234 iterations and all the obtained solutions are feasible.  Besides, the rate of global optimization reaches {\large $99.9\%$}, which is extremely high compared with the rate less than {\large $37\%$} obtained by the Hopfield neural network (Akiyama, et al., 1991; Nozawa, 1992, 1994). Finally as shown in Table \ref{t1}, $100\%$ of the rate of global optimization is also realized by simulations with $\beta =0.003$ for this toy problem with 10 cities.

On the other hand, a kind of SSA with the Metropolis algorithm is also applied to the same 10 city TSP to compare with CSA in terms of computational cost. In this calculation, SSA adopts the same cooling schedule as that of CSA, namely the exponential cooling schedule of eqn.(\ref{e5}) with $\beta =0.00004$ for annealing. The results of SSA for this 10 city TSP with 100 different initial conditions are summarized as follows:

\begin{center}
Rate of global minima : 92 \%\\
Average iterations : 58700
\end{center}

Here, one iteration means one cyclic updating of all neuron states according to the Metropolis criterion and the temperature is decreased by one step  after each iteration.

Obviously, CSA is much faster and more efficient than this SSA for the 10 city TSP.

Next, we adopt the  48 U.S.A. city TSP (Reinelt, 1994) as another example for CSA; the  parameters are set as follows:  

\[
 k=0.9; \epsilon = 1/250; I_{0}=0.5;  z(0)=0.10; \alpha =0.015, 
 \beta =5.0 \times 10^{-5}; W_{1}=1; W_{2}=1/3
\]

Fig.\ref{f91} and Fig.\ref{f92} are  the optimal route (the route length = 10628) and a near-optimal route (the route length = 10805) in the 48 U.S.A. city TSP, respectively. The results with 100 different initial conditions are summarize in Table \ref{t11}, where the average iterations is 25632. The results show that $84 \%$ solutions converge to  near-optimal routes like Fig.\ref{f92} while $5\%$ solutions converge to the optimal route of Fig.\ref{f91} from the 100 random initial conditions.

The calculations in this paper are based on cyclic updating of neuron states. Therefore, as mentioned before, one iteration means that all neuron states are cyclically updated once.  On the other hand, we have obtained similar results with a randomly asynchronous updating scheme of TCNN for the  combinatorial optimization problems (Chen \& Aihara, 1993, 1994a, 1994b), where the  difference between cyclic updating and randomly asynchronous updating is only that the periodic orbits due to the cyclic updating disappeared in the chaotic searching phase when the randomly asynchronous updating scheme was adopted.

\clearpage
\pagestyle{myheadings}
\markright{APPLICATION TO MAINTENANCE SCHEDULING FOR GENERATORS}

\section{Application to Maintenance Scheduling for Generators in a Power System}

Every generating unit in power systems should be maintained once during the specified horizon such as during one year or two years.  The maintenance scheduling for generators is to level the reliability risks with respect to the load by properly maintaining  each generating unit throughout the horizon.  Generally the reserve margin which is defined by the rate of generating capacity reserve versus load as shown in Fig.\ref{f10}, is widely used as a risk index (Chen \& Toyoda, 1991). In Fig.\ref{f10}, the vertical and horizontal lines for each unit block indicate its generation capacity and maintenance period, respectively.  Since generating units have different sizes of capacities and different maintenance periods, to level the reserve margins throughout the horizon becomes a  large-scale combinatorial optimization problem. Besides, there are two practical constraints which should be taken into consideration; 
the one is that each unit has to be maintained during its specified periods, and the other is that units belonging to the same power plant should not be maintained simultaneously in the same period.

\subsection{Formulation of maintenance scheduling for generators}

In order to level the reserve margins, the maintenance scheduling for generators can be formulated as the following 0-1 integer programming (Chen, 1993):

\begin{equation}
\mathop{Max}_{x_{ij}} \ \mathop{Min}_{j=1,...,h} R_{j}= \frac{\sum_{i=1}^{n} G_{i}(1- \sum_{k=j-mp_{i}+1}^{j}x_{ik})-D_{j}}{D_{j}}  \label{e20} 
\end{equation}

\begin{eqnarray}
subject \ \ to \ \ \  & \sum_{k=hf_{i}}^{ht_{i}}x_{ik}=1 \ \ \ (i=1,...,n)
\label{e21} \\
 & (\sum_{\stackrel{\scriptstyle i \in A_{l}}
{k=j-mp_{i}+1}}^{j}x_{ik})
(\sum_{\stackrel{\scriptstyle m \in A_{l}}{m \neq i}}
\sum_{k=j-mp_{m}+1}^{j}x_{mk}) = 0 \nonumber \\
       &  (i=1,...,n;j=1,...,h)                  \label{e22} 
\end{eqnarray}

where, $x_{ij}$ represents whether the unit-$i$ is maintained for $mp_{i}$ periods from period-$j$ or not, i.e.,

\[ x_{ij}=  \left\{ \begin{array}{ll} 1 & \mbox{if unit}-i \ \mbox{is maintained for} \ mp_{i} \ \mbox{periods from period}-j  
              \\ 0 & \mbox{otherwise}
	      \end{array} \right.\]

$G_{i}$: generating capacity of unit-$i$, $n$: total number of units, $h$:total periods for the horizon, $mp_{i}$: maintenance periods of unit-$i$, $D_{j}$: load in period-$j$, $R_{j}$: reserve margin in period-$j$, $hf_{i} \sim ht_{i}$: specified maintenance periods for unit-$i$, $A_{l}$: set for units belonging to power plant-$l$.

Eqn.(\ref{e21}) represents the constraint that each unit should be maintained once and only once during its specified periods. Eqn.(\ref{e22}) is the constraint which represents that the units belonging to the same power plant should not be maintained simultaneously in the same period.

In the left-hand side of eqn.(\ref{e22}), the first term calculates whether the unit-$i$ is maintained in period-$j$ or not, namely, the first term is larger than zero if maintaining, otherwise this term is zero. The second term checks whether the other units belonging to the same power plant as unit-$i$ are maintained  in the same period or not.

>From the definition of $x_{ij}$, $(hf_{i}-ht_{i}+1)$ neurons are necessary for one unit.  Therefore, the total number of neurons is $\sum_{i=1}^{n}(hf_{i}-ht_{i}+1)$ for the  neural network computing the maintenance scheduling.

According to Eqns.(\ref{e20})-(\ref{e22}), the energy function $E$ for minimization is constituted as follows: 

\begin{eqnarray}
 E=\sum_{j=1}^{h}-\lambda_{j}R_{j}^{2}+ \nonumber  
 \sum_{i=1}^{n}\frac{W_{1}}{2}(1-\sum_{k=hf_{i}}^{ht_{i}}x_{ik})^{2} +\\ 
 \sum_{i=1}^{n}\sum_{j=1}^{h}\frac{W_{2}}{2}(\sum_{\stackrel{\scriptstyle i \in A_{l}}{k=j-mp_{i}+1}}^{j}x_{ik})(\sum_{\stackrel{\scriptstyle m \in A_{l}}{m \neq i}}\sum_{k=j-mp_{m}+1}^{j}x_{mk})  \label{e23}
   \end{eqnarray}

where, the first term in the right-hand side is weighted summation for the total reserve margins; the second and third terms correspond to the constraints;  $W_{1},W_{2}$
 are the coupling strengths corresponding to the constraints; $\lambda_{j}$ is a positive weighting parameter.
 
 It is evident that minimizing eqn.(\ref{e23}) is not for leveling the reserve margins, but for maximizing the weighted summation of the total reserve margins meanwhile satisfying the constraints. 
 Therefore, in order to level the reserve margins approximately, the weighted parameters $\lambda_{j}$ should be set appropriately for the concerned problems (Chen, 1993).

Substituting $-\partial E/\partial x_{ij}$ into eqns.(\ref{e3})-(\ref{e5}), we obtain the transiently chaotic neural network for the maintenance scheduling of generators.

\subsection{Simulation of maintenance scheduling}

We apply the transiently chaotic neural network to the maintenance scheduling for generators in a practical power system with 117 units belonging 40 power plants (Chen \& Aihara, 1993). The horizon (one year) is divided into 26 periods (1 period is 2-week); $\lambda_{j}=W1=W2=1$; the total number of neurons is  2384.

All the solutions obtained by simulations with 100 random initial conditions are feasible and the minimal reserve margins among 26 periods have the same value of 24.2\%.  It has been observed that the solutions 
 largely depend on the setting of the weighting parameter $\lambda_{j}$ compared with other parameters $W_{1}$ and $W_{2}$. 
 The average CPU time for one solution is approximately 26 seconds with  FACOM M-770 (22MIPS).

\section{Discussions}

To cope with the combinatorial optimization problems, we have proposed the 
chaotic simulated annealing with the 
transiently chaotic neural network, as an approximation method. The transiently chaotic dynamics is utilized for global searching and self-organizing where accumulation of refractory or self-inhibitory effects of chaotic neuron models prevents the network from getting stuck at local minima. After the transiently chaotic dynamics has vanished, the proposed neural network then is fundamentally reined by the gradient descent dynamics and usually converges to a stable equilibrium point like the Hopfield neural network.  The features of the present method are summarized as follows:

\begin{enumerate}
\item The dynamics of the transiently chaotic neural network is characterized as  transient chaos.
\item The proposed model gradually approaches, through the transient chaos, to dynamical structure similar to the Hopfield neural network which  converges to a stable equilibrium point.
\item In contrast to the conventional {\it stochastic simulated annealing}, the  mechanics of the proposed method is regarded  as  {\it chaotic simulated annealing}.
\item Routes of optimization by the chaotic simulated annealing are through successive  bifurcations from strange attractors to an equilibrium point.

\item The proposed annealing mechanics is realized as damping of negative (inhibitory) self-feedback connection weights.
\end{enumerate}

The main difference between SSA and CSA is that the annealing process of CSA is associated with a series of the  bifurcations, rather than decreasing  statistic fluctuations based on the Monte Carlo scheme in SSA. Since the mechanics of CSA is based on autonomous searching in a reduced possibly fractal subspace by simple deterministic dynamics, CSA can be expected to require just short computation time. The characteristics of CSA and SSA are summarized in Table \ref{t2}.   

This method of CSA provides a new mechanism of deterministic simulated annealing (DSA), which is different from other methods of DSA, such as gain sharpening of the Hopfield network (Hopfield \& Tank, 1985; Lee \& Sheu, 1991; Akiyama et al., 1991)  and mean field approximation annealing (Peterson \& Anderson, 1987; Hinton, 1989; Geiger \& Girosi, 1991, Masui \& Matsuba, 1991; Peterson \& Soderberg, 1993).        

Numerical experiments of TCNN on TSP and the maintenance scheduling problem for generators have shown that the present approach has high efficiency to converge to globally optimal or near-optimal solutions on these small-scale problems of combinatorial optimization. 
When compared with SSA, CSA has a peculiar characteristic that the searching region in the state space is restricted to a small possibly fractal structure whose volume should be zero with respect to the Lebesgue measure of the whole state space.  The possible reason why CSA has the high ability of global search in spite of its small searching region is that the deterministic dynamics due to mutual interactions among neurons, which reflect the problem structure such as costs and constraints,  restricts the searching region efficiently so that the fractal region includes at least some parts of the basins associated with  globally optimal or near-optimal solutions. Besides,to attain faster speed, hardware
implementations of CSA (Horio et al.,1994) is also an important subject.

One may ask what is difference between  stochastic noises and  chaotic noises if the stochastic noises are generated by a pseudo-random number generator in digital computers. 
There are several aspects on this problem. Firstly, the usual chaotic noises are different from the conventional pseudo-random noises produced by say the linear congruence method  because the former own short-term autocorrelation peculiar to the chaotic dynamics while the latter are nearly uncorrelated (Hayakawa \& Sawada, 1994). 
In other words, although the pseudo-random noises are generated by a kind of chaotic dynamics, they are particularly designed to be almost uncorrelated due to 
 the largely positive Lyapunov exponent with homogeneous expansion and coarse-graining to integers. It is suggested by comparing effects of three kind of additive noises to gradient descent dynamics for optimization, namely  pseudo-random noises,  chaotic noises by the logistic map (May, 1976) and their randomly shuffled version,  that the short-term correlation of the logistic chaos is useful for optimization (Hayakawa \& Sawada, 1994). Another aspect is mutual interaction among chaotic elements realized in TCNN. It is reported by globally coupled maps (Kanako, 1992) that coupled chaotic dynamics can generate inherent characteristics like breakdown of the law of large numbers and coherent irregularity (Tomita, 1984), which cannot be produced by just coupling of random noises.

Finally, we should stress that, although the proposed approach to combinatorial optimization problems by transiently chaotic neural networks with chaotic simulated annealing provides a new kind of approximation methods, further studies, in particular, strict comparison with the excellent techniques already in use (Johnson, 1987) are indispensable for the evaluation of the true ability. 
 In fact, it seems that the TCNN with CSA can not compete with a heuristics  special for a certain problem, such as the Lin-Kernighan algorithm for TSP. As a not problem-specific but general method, however, CSA may be worth examing the ability for combinatorial optimization.

It should be also noted, on the other hand, that although the number of cities of the TSP analysed in this paper is so small as 10 and 48, the dimension of the corresponding chaotic mapping are 100 and 2304 which numbers are fairly large as the dimension of chaotic dynamical systems so far studied.  The analyses on the nonlinear dynamics of such large-scale chaotic systems and its relationships with such information processing  as combinatorial optimization are important and interesting future problems.

\clearpage
\pagestyle{myheadings}
\markright{Captions of Figure and Table}

\begin{center}
Captions of Figure and Table
\end{center}

\vspace*{1cm}

\begin{itemize}
\item Figure \ref{f1} : Time evolutions of $x(t), z(t)$ and the Lyapunov exponent $\lambda$ in the single neuron dynamics. $x(t)$: the output of the neuron, $z(t)$: the self-feedback connection weight, or the damping variable corresponding to the temperature in annealing process.
\item Figure \ref{f2} : Time evolutions of the single neuron model with different values of $\beta$.
 (a) $\beta =0.002, \gamma =0.0$, (b) $\beta =0.0008, \gamma =0.0$.
\item Figure \ref{f4} : The Hopfield-Tank original data of TSP with  10 cities.
\item Figure \ref{f5} : Time evolutions of continuous and discrete energy functions and $z(t)$ in simulation of TCNN for TSP with 4 cities. $\beta =0.001, \alpha =0.015$. 
\item Figure \ref{f6} : Time evolution  of each neuron output $x_{ij}$ for 
TSP with 4 cities. $\alpha =0.015, \beta =0.001$.
\item Figure \ref{f7} : Amplification of $x_{11}$  in Fig.5.
\item Figure \ref{f8} : Time evolutions of the continuous energy function with different values of $\beta$. (a) $\beta =0.0001, \alpha =0.015$, (b) $\beta =0.1, \alpha =0.015$.
\item Figure \ref{f9} : Time evolutions of the continuous energy function with different values of $\alpha$. (a) $\beta =0.001, \alpha =0.001$, (b) $\beta =0.001, \alpha =0.1$.
\item Figure \ref{f91} : The optimal route for the 48 U.S.A. City TSP 
\item Figure \ref{f92} : A near-optimal route for the 48 U.S.A. City TSP 
\item Figure \ref{f10} : Maintenance scheduling for generators in a power system.
\item Table \ref{t1} : Results of 5000 different initial conditions  for each value of $\beta$ on the 10 city TSP by CSA.

\item Table \ref{t11} : Results of 100 different initial conditions on the 48 U.S.A. city TSP by CSA. 
\item Table \ref{t2} : Comparisons between the chaotic simulated annealing (CSA) and the conventional stochastic simulated annealing (SSA) in neural networks.
\end{itemize}

\clearpage
\pagestyle{myheadings}
\markright{Figure and Table}

\begin{figure}
\vspace*{8.5cm}
\begin{center}
$(\beta =0.001; \gamma =0.0)$\\
 \end{center}
\caption{Time evolutions of $x(t), z(t)$ and the Lyapunov exponent $\lambda$ in the single neuron dynamics}
\label{f1}
\end{figure}
   
\clearpage
   
\begin{figure}
   \vspace*{5cm}
\begin{center}
(a) \ \ \ $(\beta =0.002; \gamma =0.0)$
\end{center}
   \vspace*{5cm}
\begin{center}
(b) \ \ \ $(\beta =0.0008; \gamma =0.0)$\\
\end{center}
\caption{Time evolutions of the single neuron model with different values of $\beta$}
   \label{f2}
\end{figure} 

\clearpage


\begin{figure}
\vspace*{6cm}
\caption{The Hopfield-Tank original data of TSP with  10 cities}
\label{f4}
\end{figure}

\clearpage

\begin{figure}
\vspace*{6.5cm}
\begin{center}
$(\beta =0.001; \alpha =0.015)$\\
\end{center}
\caption{Time evolutions of continuous and discrete energy functions and $z(t)$ in simulation of TCNN for TSP with 4 cities}
\label{f5}
\end{figure}

\clearpage

\begin{figure}
\vspace*{8cm}
\begin{center}
$(\beta =0.001; \alpha =0.015$)\\
\end{center}
\caption{Time evolution  of each neuron output $x_{ij}$ for
TSP with 4 cities}
\label{f6}
\end{figure}

\clearpage

\begin{figure}
\vspace*{3.0cm}
\begin{center}
$(\beta =0.001; \alpha =0.015)$\\
\end{center}
\caption{Amplification of $x_{11}$  in Fig.5}
\label{f7}
\end{figure}

\clearpage

\begin{figure}
\vspace*{2.5cm}
\begin{center}
(a) \ \ \ $(\beta =0.0001; \alpha =0.015)$
\end{center}
\vspace*{2.5cm}
\begin{center}
(b) \ \ \ $(\beta =0.1; \alpha =0.015)$\\
\end{center}
\caption{Time evolutions of the continuous energy function with different values of $\beta$}
\label{f8}
\end{figure}

\clearpage

\begin{figure}
\vspace*{2.5cm}
\begin{center}
(a) \ \ \ $(\beta =0.001; \alpha =0.001)$
\end{center}
\vspace*{2.5cm}
\begin{center}
(b) \ \ \ $(\beta =0.001; \alpha =0.1)$\\
\end{center}
\caption{Time evolutions of the continuous energy function with different values of $\alpha$}
\label{f9}
\end{figure}

\clearpage

\begin{figure}
\vspace*{6cm}
\begin{center}
(Tour Length = 10628)
\end{center}
\caption{The optimal route for the 48 U.S.A. city TSP}
\label{f91}
\end{figure}

\clearpage

\begin{figure}
\vspace*{6cm}
\begin{center}
(Tour Length = 10805)
\end{center}
\caption{A near-optimal route for the 48 U.S.A. city TSP}
\label{f92}
\end{figure}

\clearpage

\begin{figure}

\vspace*{11cm}

\caption{Maintenance scheduling for generators in a power system}
\label{f10}
\end{figure}

\clearpage

\begin{table}

\caption{Results of 5000 different initial conditions  for each value of $\beta$ on the 10 city TSP by CSA}
\label{t1}
\vspace*{5mm}

\begin{tabular}{c|r|r|r|r}\hline 
                  &                &                &                &                \\
$(\alpha =0.015)$ & $\beta =0.015$ & $\beta =0.010$ & $\beta =0.005$ & $\beta =0.003$ \\ 
                  &                &                &                &                \\ \hline
                  &                &                &                &                \\
Rate of global minima (\%) & 4946 (98.9\%) & 4969 (99.4\%) & 4998 (99.9\%) & 5000 (100\%) \\ 
                  &                &                &                &                \\ \hline
                  &                &                &                &                \\
Rate of local minima \ (\%) & 31 (0.6\%)   & 13 (0.3\%) &  2 (0.1\%) & 0 (0.0\%) \\ 
		  &                &                &                &                \\ \hline
                  &                &                &                &                \\ 
Rate of infeasible solutions (\%) & 23 (0.5\%)   & 18 (0.3\%)    &  0 (0.0\%) & 0 (0.0\%)      \\ 
		  &                &                &                &                \\ \hline
                  &                &                &                &                \\
Average iterations &               &               &       &        \\
for convergence   & 81            &  119          &    234  & 398   \\ 
		  &                &                &                &                \\ \hline
\end{tabular}
\end{table}

\clearpage

\begin{table}

\caption{Results of 100 different initial conditions on the 48 U.S.A. city TSP by CSA}
 \label{t11}
 \vspace*{5mm}

 \begin{center}
 \begin{tabular}{r|r|r}\hline
	      &             &           \\
Route length & Number (\%) & Error \%  \\
	      &             &           \\ \hline
              &             &           \\
${}^{*}$10628 & 5 (5\%)     & 0\%       \\
	      &             &           \\ \hline
              &             &           \\
10805         & 84 (84\%)   & 1.665\%   \\
	      &             &           \\ \hline
              &             &           \\
10937         & 3 (3\%)     & 2.907\%   \\
	      &             &           \\ \hline
              &             &           \\
10992         & 3 (3\%)     & 3.425\%   \\
	      &             &           \\ \hline
              &             &           \\
others        & 5 (5\%)     & --------- \\
	      &             &           \\ \hline
  \end{tabular}
  \vspace*{5mm}
  
${}^{*}$10628 : the best length\\
Error : error to the best length
 \end{center}
  \end{table}
  
\clearpage

\begin{table}

\caption{Comparisons between the chaotic simulated annealing (CSA) and the conventional stochastic simulated annealing (SSA) in neural networks}
\label{t2}
\vspace*{5mm}

\begin{tabular}{c|c|c}\hline
                  &                       &                           \\
                  & {\bf CSA}             & {\bf SSA}                 \\
	          &                       &                           \\ \hline
                  &                       &                           \\
Searching mechanics & deterministic chaos & stochastic transition      \\
		  &                       &                           \\ \hline  
                  &                       &                           \\
		  & bifurcations to a near- & realization of a near-   \\ 
Annealing and Convergence & optimal equilibrium point  & optimal ground state \\  
& by decreasing self-connection & with decreasing temperature \\
		  &                       &                           \\ \hline
                   &                       &                           \\
Global searching space & fractal subspace & all possible states       \\
		  &                       &                           \\ \hline 
                   &                       &                           \\
Local searching  &  uniquely             & probabilistically          \\ 
direction         & determined by mutual  & determined by mutual       \\ 
             & interactions among neurons & interactions among neurons \\
                   &                       &                           \\  \hline
		   \end{tabular}
\end{table}

\end{document}